\documentclass{camera}
\usepackage{graphicx}  

\def\apj{\emph{ApJ.}}
\def\apjl{\emph{ApJ. Lett.}}

\begin{document}
\pagenumbering{arabic}

%
\title{The lag-luminosity relation in the gamma-ray burst source-frame}

%
\author{T. N. Ukwatta$^{1,2}$, K. S. Dhuga$^1$, M. Stamatikos$^{2,3}$, T.
Sakamoto$^2$, W. C. Parke$^1$, S. D. Barthelmy$^2$ \and N.
Gehrels$^2$}

%
\organization{$^1$The George Washington University, 725 21st St,
NW, Washington, D.C. 20052, USA. \\ $^2$NASA Goddard Space Flight
Center, Astroparticle Physics Laboratory, Code 661, Greenbelt, MD
20771, USA.\\ $^3$Center for Cosmology and Astro-Particle Physics
(CCAPP) Fellow/Department of Physics, The Ohio State University,
191 West Woodruff Avenue, Columbus, OH 43210, USA.}

\maketitle

\begin{abstract}
We have investigated the empirical lag-luminosity relation in the
Gamma-ray Burst (GRB) source-frame. We selected two energy bands
($100-200$ keV and $300-400$ keV) in the GRB source-frame, which
after redshift correction, lie in the observer-frame energy range
of the $Swift$ Burst Alert Telescope (BAT). The spectral lags
between these energy channels are then presented as a function of
the isotropic peak luminosity of the GRBs in the sample.
\end{abstract}

%
\section{Introduction}
Spectral lag is a common feature in Gamma-ray Bursts (GRBs). The
lag is defined as the difference in time of arrival of high and
low energy photons and is considered positive when the high-energy
photons arrive earlier than the low energy ones. Norris et al.
reported a correlation between spectral lag and the isotropic peak
luminosity of GRBs based on a limited sample~\cite{norris2000}.
Subsequently, various authors have studied the lag-luminosity
relation using arbitrary observer-frame energy bands of various
instruments~\cite{Ukwatta2009lag,Hakkila2008}.

\begin{figure}[htp]
\centering
\includegraphics[width=12cm, angle=0]{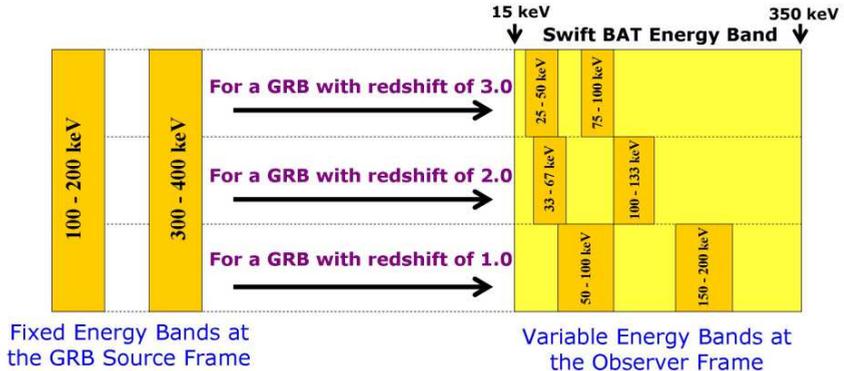}
\caption{Fixed energy bands at the GRB source-frame are projected
to various energy bands at the observer-frame, depending on the
redshift.}
\label{fig01} 
\end{figure}

Typically the spectral lag ($\tau$) is extracted in two arbitrary
energy bands in the observer-frame. However, because of the
redshift (z) dependance of GRBs, the two energy bands can
correspond to multiple energy bands in the source-frame thus
introducing a variable energy dependant factor, which is difficult
to take into account. We avoid this difficulty by defining two
energy bands ($100-200$ keV and $300-400$ keV) in the GRB
source-frame and projecting these two bands into the
observer-frame using the relation $E_{\rm observer}=E_{\rm
source}/(1+z)$. For our sample of GRBs, after projecting to the
observer-frame, the selected energy bands lie in the $Swift$ Burst
Alert Telescope (BAT)~\cite{Gehrels2004} energy range ($15-350$
keV; see Fig.~\ref{fig01}). Then we extracted the spectral lags
between these energy channels, correct them for cosmological time
dilations using the relation, $\tau_{\rm source}=\tau_{\rm
observer}/(1+z)$ and plot them as a function of the isotropic peak
luminosity.

\section{Methodology}

$Swift$ BAT is a highly sensitive coded aperture
instrument~\cite{barthelmy2005}. BAT uses the modulation pattern
resulting from the coded mask to facilitate background subtraction
in the light curve and the spectral processing. This process is
called mask weighting. We used the mask weighted,
background-subtracted light curves, as well as non-mask weighted
light  curves in our analysis. The non-mask weighted light curves
have  about 30\% better signal-to-noise ratio than the mask
weighted light curves. Hence, when the signal-to-noise ratio of
mask weighted light curves was too small, non-mask weighted light
curves were used to extract spectral lags.

\begin{figure}[htp]
\centering
\includegraphics[width=12cm, angle=0]{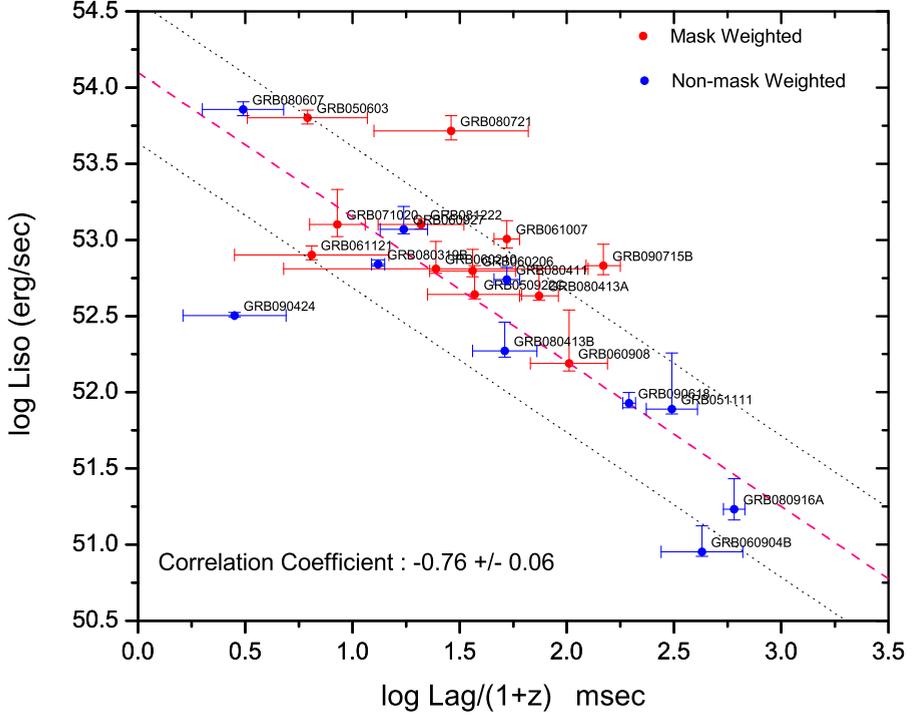}
\caption{The spectral lags between the source-frame energy range
bands $100-200$ keV and $300-400$ keV are plotted against the
isotropic peak luminosity. Red data points indicate spectral lag
values obtained from mask weighted light curves and blue indicate
lag values extracted from non-mask weighted light curves.}
\label{fig02} 
\end{figure}

The spectral lags were extracted using improved cross-correlation
function (CCF) analysis method described in
reference~\cite{Ukwatta2009lag}. In this method the spectral lag
is extracted by finding the time delay corresponding to the global
maximum of the CCF. A Gaussian curve was fitted to the CCF to
extract the spectral lag. The uncertainty in the spectral lag is
obtained by simulating 1,000 light curves and repeating the
aforementioned procedure. The standard deviation of the simulated
spectral lag values is taken as the lag uncertainty. The isotropic
peak luminosity ($L_{\rm iso}$), for each GRB, is obtained by
calculating the peak flux for the source-frame energy range, 1.0
keV to 10,000 keV, using the observed spectral-fit parameters and
the luminosity distance as described in ~\cite{Ukwatta2009lag}.

\section{Results}

We used the sample of 31 GRBs in reference~\cite{Ukwatta2009lag}.
However, some of the GRBs in the sample did not have enough
signal-to-noise to extract a spectral lag limiting the final
sample to 22 GRBs.

In our analysis, we have selected the GRB source-frame energy
bands $100-200$ keV and $300-400$ keV due to the energy range of
$Swift$ BAT and the redshift range of our sample. For mask
weighted light curves, the effective energy range of the BAT
reduces to 15-200 keV, which implies a GRB redshift range of 1 to
5.7. However, for the non-mask weighted light curves we don't have
such a restriction and the full BAT energy range ($15-350$ keV)
can be utilized. In this case the redshift lower limit can be as
low as 0.2. Our final sample of 22 GRBs have redshift ranging from
0.54 to 5.46.

In Fig.~\ref{fig02}, $L_{\rm iso}$ is plotted as a function of
redshift corrected spectral lag. The spectral lag is
anti-correlated with $L_{\rm iso}$ with a correlation coefficient
of -0.76 $\pm$ 0.06. The index of the best-fit power-law is $-0.9
\pm 0.1$, which is shallower compared to observer-frame results
from Norris et al. 2000 \cite{norris2000} ($\sim \, -1.14$) and
Ukwatta et al. 2009 \cite{Ukwatta2009lag} ($\sim -1.4$).

\section{Summary and Discussion}

We have investigated the spectral lag between $400-300$ keV and
$200-100$ keV energy bands at the GRB source-frame by projecting
these bands to the observer-frame. This is a step forward in the
investigation of lag-luminosity relation since all previous
investigations used arbitrary observer-frame energy bands. The
correlation coefficient of $\sim \, -0.76$ shows a significant
improvement over the average correlation coefficient of $\sim \,
-0.68$ reported in the reference \cite{Ukwatta2009lag}.



%
\end{document}